%% file: 2EP.tex
\documentclass[aps,prl,superscriptaddress,twocolumn,showpacs]{revtex4}
\usepackage{amsmath}
\usepackage{graphicx}
\usepackage{amssymb}
\usepackage{bbm, color, soul}
\usepackage{graphicx}
\usepackage{adjustbox}

\newcommand*{\tn}[1]{{\textnormal{#1}}}

\usepackage{color}

 \newcommand{\fRP}{\mathsf{RP}} 

\usepackage{tikz}

\usepackage{amsthm}
\usepackage{tikz-cd}
\usepackage[all]{xy}
\usepackage{amsfonts}
\usepackage{mathrsfs}
\usepackage{hyperref}
\usepackage{MnSymbol} 

\input{header_labeling.tex}

\input{header_newcommand.tex}

\begin{document}

\title[]{Topological Switching via Exceptional Point Pairs in an Optical Microcavity}

\author{Kyu-Won \surname{Park}}
\email{parkkw7777@gmail.com}
\affiliation{Department of Mathematics and Research Institute for Basic Sciences, Kyung Hee University, Seoul, 02447, Korea}

\author{KyeongRo \surname{Kim}}
\email{kyeongrokim14@gmail.com}
\affiliation{Research Institute of Mathematics, Seoul National University, Seoul 08826, Korea}

\author{Jinuk \surname{Kim}}
\affiliation{Korea Research Institute of Standards and Science, Daejeon 34113, Korea}

\author{Muhan  \surname{Choi}}
\affiliation{Digital Technology Research Center, Kyungpook National University, Daegu 41566, Korea}

\author{Kabgyun \surname{Jeong}}
\email{kgjeong6@snu.ac.kr}
\affiliation{Research Institute of Mathematics, Seoul National University, Seoul 08826, Korea}
\affiliation{School of Computational Sciences, Korea Institute for Advanced Study, Seoul 02455, Korea}

\date{\today}

\begin{abstract} Conventional mode switching mechanisms, which rely on dynamically encircling exceptional points (\textsf{EP}s) through non-adiabatic transitions (NATs), suffer from intrinsic nonlinear dynamics that hinder precise control and reproducibility in experimental settings. Additionally, these methods exhibit low transmission efficiencies due to path-dependent attenuation, limiting their effectiveness in optical switching and logic gate applications.
To overcome these limitations, we propose a novel mode switching approach that leverages a pair of \textsf{EP} configuration in an optical microcavity, characterized by superradiance and subradiance. This mechanism exploits the topological structure of the Riemann surface to enable robust mode switching control and tunable Q-factor through purely adiabatic encircling. Furthermore, topological protection validated via braid isotopy ensures robustness against noise and parametric perturbations, facilitating a compact, robust, and adaptive non-Hermitian system.

\end{abstract}

\maketitle

\section{INTRODUCTION}
Exceptional points (\textsf{EP}s) arise in non-Hermitian systems where eigenvalues and eigenvectors coalesce, leading to exotic physical phenomena and novel device functionalities~\cite{R09, IJ15, W04, T66}. These unique singularities have been extensively studied in various physical platforms, including carbon nanotubes~\cite{GL18}, nano-optomechanical systems~\cite{LP18, WZ21}, photonic crystals~\cite{ZP18, CY21}, and optical microcavities~\cite{WS17, CW21}. \textsf{EP}s are also associated with various intriguing phenomena, including parity-time symmetry~\cite{RK18, LT21}, phase transitions~\cite{MY16, AR21}, chirality~\cite{TG18, HG22}, Shannon entropy~\cite{KJ18}, and topological braiding of eigenvalues~\cite{Guria2024, Wang2021, Guo2023, Guo2021, Rao2024}. Especially, the topological braiding of eigenvalues in complex energy spectra has emerged as a key concept in non-Hermitian physics, playing a fundamental role in topological classification and topologically governed mode transitions. Furthermore, \textsf{EP}s have been utilized in ultra-sensitive sensing~\cite{WS17, J14, RJ22}, high-precision gyroscopes~\cite{MM18, MA19}, and non-reciprocal photonic devices~\cite{Feng2013}, leveraging their exceptional spectral properties.

In particular, the unique topological properties of \textsf{EP}s, governed by the Riemann surface structure of the complex energy spectrum, play a crucial role in several key phenomena. These include stroboscopic (i.e., adiabatic) encircling~\cite{Zhong18, Dembowski01, Gao15, Ding16}, mode switching~\cite{DM16, XA22, Zhang2019, Khurgin2021, Liu2021}, and topological energy transfer~\cite{Schumer2022, Wang20211, Xu2016}.
Notably, stroboscopic encircling of an \textsf{EP} in two-level systems primarily leads to mode exchange and geometric phase accumulation but generally does not induce mode switching~\cite{R09, IJ15, W04, Lee10, Nesterov08}. In contrast, dynamically encircling an \textsf{EP} in such systems, which involves non-adiabatic transitions (NATs), enables mode switching. Particularly, this asymmetric mode switching mechanism has significantly advanced the development of optical switching~\cite{Li2020, Li2024}, logic gates~\cite{Yang2022, Zhang2023}, and other photonic applications~\cite{Yu2021, Longstaff2019}.
However, despite its utility, NATs exhibit inherently nonlinear dynamics, making precise control and reproducibility in experimental settings highly challenging~\cite{DM16, Berry11, BerryUzdin11}. Additionally, path-dependent losses further reduce transmission efficiency, complicating practical implementations~\cite{Zhang2019, Zhang18, Li20}. Moreover, eigenvalue discontinuities arising in NATs compromise the robustness of conventional systems. Therefore, achieving reliable mode control requires a more robust, topologically protected approach.
\setlength{\parskip}{0pt}  
\setlength{\parindent}{10pt} 

In this work, we propose a topological switching scheme utilizing a pair of-\textsf{EP}s configuration in an optical microcavity, where superradiance significantly enhances, whereas subradiance significantly suppresses decay rates~\cite{HI13, HI14}. Our approach utilizes braid topology on a Riemann surface for robust mode switching, ensuring deterministic control. This mechanism enables topologically protected tuning of laser frequency and output power, ensuring robustness against parametric perturbations. Conventional compact CW laser control methods, such as acousto-optic modulators (AOMs)~\cite{Liu23, He21} and variable optical attenuators (VOAs)~\cite{Xiang23, Mourikis24}, are constrained by size limitations and integration challenges, making them less suitable for miniaturized photonic systems. Meanwhile, direct modulation techniques-including injection current tuning~\cite{Russer82, Yamaoka21}, pump power control~\cite{Wang20, He23}, and temperature regulation~\cite{Huang24, Lai22}-suffer from thermal instability, precision demands, and noise sensitivity. Our \textsf{EP}-based approach mitigates these limitations, enabling robust output tuning to parametric variations. This makes it particularly suitable for compact laser systems requiring high precision, such as medical lasers~\cite{MN20, Schomacker1990}, miniaturized LiDAR~\cite{HuangX24, Jeong24}, atomic clocks~\cite{Zheng24, Hinkley13}, and optical tweezers~\cite{Bustamante21, Grun24}.

\section{Non-Hermitian Hamiltonians and Optical Microcavities}
Non-Hermitian effects naturally arise in open physical systems as a result of interactions with their environment. A non-Hermitian Hamiltonian describing these phenomena takes the form $H_{\rm NH} = H_S + V_{SE} G_E^{(\rightarrow)} V_{ES}$, where $H_S$ is the Hermitian Hamiltonian of the closed system, $G_E^{(\rightarrow)}$ is the outgoing Green function of the environment, and $V_{SE}$ ($V_{ES}$) represents the system-environment coupling~\cite{R09, IJ15}.

Assuming two dominant interacting eigenstates, $H_{\rm NH}$ can be expressed as
\begin{align}
H_{\rm NH} =
\begin{pmatrix}
\lambda^{1} & g \\
g & \lambda^{2}
\end{pmatrix},
\label{eq2}
\end{align}
where $\lambda^{i} \in \mathbb{C}$ are self-energy terms, and $g \in \mathbb{C}$ represents the coupling coefficient that incorporates both Hermitian and non-Hermitian contributions. The eigenvalues of this system are given by $\lambda_{\pm} = \frac{\lambda^{1} + \lambda^{2}}{2} \pm \eta$, where $\eta = \sqrt{\frac{(\lambda^{1} - \lambda^{2})^2}{4} + g^{2}}$ represents the eigenvalue splitting.

Conventional \textsf{EP} models assume a predominantly Hermitian coupling ($g_{12} = g_{21}^{*}$, $\text{Re}(g) > \text{Im}(g)$), leading to Riemann surfaces of the form $f(z) = z^{1/N}$ for $N =2, 3, 4$. In contrast, we extend this framework to a non-Hermitian coupling regime where $g_{12} \neq g_{21}^{*}$ and $\text{Im}(g) > \text{Re}(g)$. This extension introduces novel topological structures in the Riemann surface, leading to new spectral features that are absent in Hermitian coupling.

\begin{figure*}
\centering
\includegraphics[width=16.0cm]{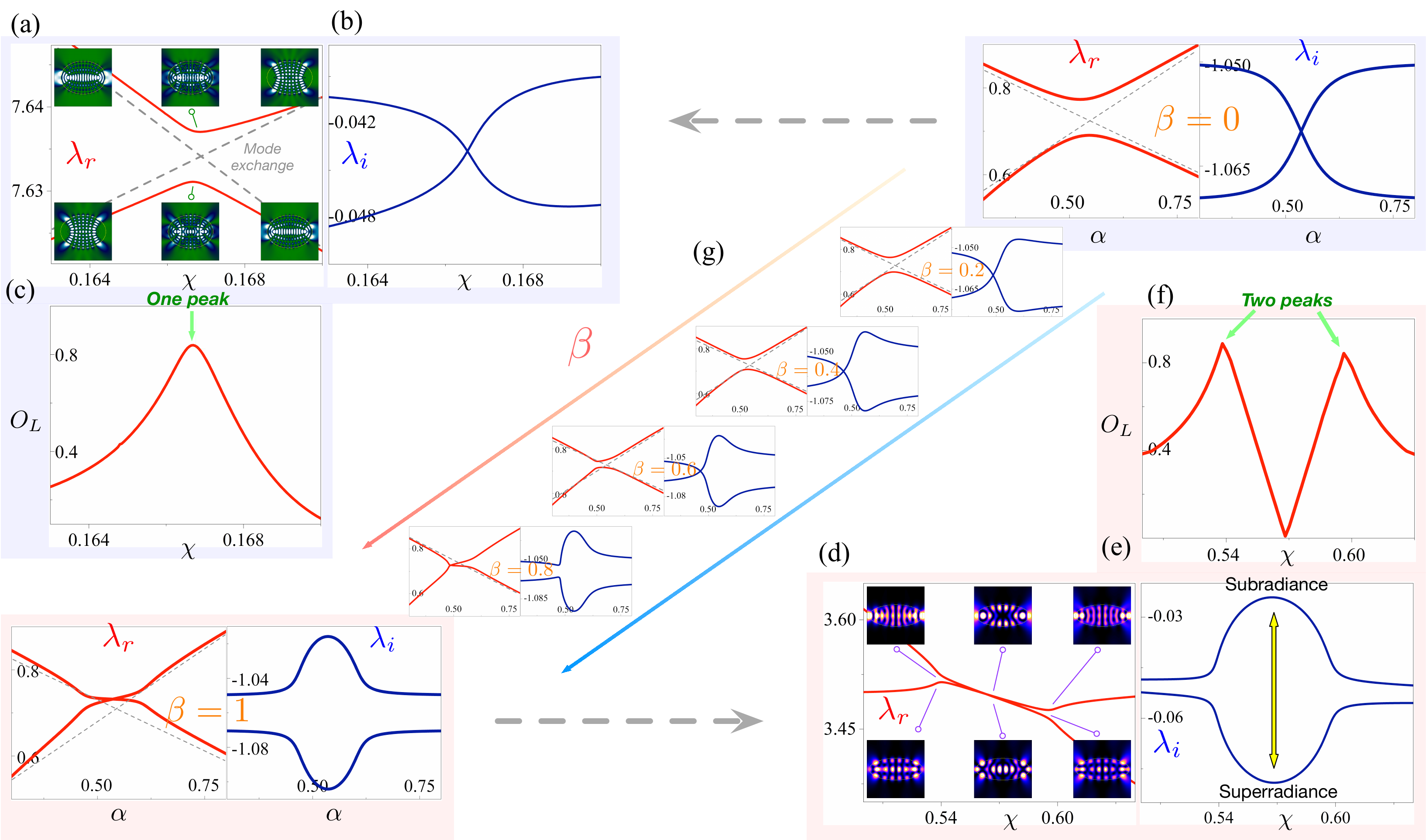}
\caption{Comparison of Landau-Zener avoided crossing (top left) and decay (width) bifurcation (bottom right). The insets show the eigenstate intensity distributions. (a, b) Avoided crossing in real eigenvalues and mode crossing in imaginary eigenvalues. (c) Overlap integral peaks at the center of the avoided crossing. (d) The real eigenvalues exhibit a crossing. (e) The imaginary eigenvalues undergo bifurcation near two critical points. (f) The overlap integral of the two interacting eigenstates peaks near the edges of the bifurcations. (g) shows how increasing $\beta$ from 0 to 1 continuously transforms the eigenvalue structure. The transition from Landau-Zener interaction to width bifurcation occurs as the imaginary component of $g$ increases, highlighting the impact of non-Hermitian coupling.}
\label{Figure-1}
\end{figure*}

Elliptical optical microcavities provide an ideal platform for investigating openness-induced effects in non-Hermitian physics. In a closed (Hermitian) system, an elliptical cavity behaves as an integrable system, where eigenvalue crossings occur without avoided crossings, resulting in Poisson eigenvalue distributions~\cite{H99, F10}. However, when openness is introduced, avoided crossings emerge as a direct consequence of non-Hermiticity. This distinction allows elliptical microcavities to serve as a controlled platform where openness effects can be systematically studied in isolation. Previous simulations confirm that avoided crossings do not appear in closed elliptical cavities but emerge exclusively in open systems~\cite{KS18, KJ18}.

Our proposed system is a two-dimensional elliptical optical microcavity with major and minor axes defined as $a = 1 + \chi$ and $b = \frac{1}{1 + \chi}$, ensuring a constant area of $\pi$. The external refractive index is fixed at $n_{\rm out} = 1$ (vacuum), while the cavity refractive index ($n_{\rm in}$) is varied alongside $\chi$ to determine \textsf{EP} locations. To compute the eigenvalues and eigenfunctions, we solve the Helmholtz equation $\nabla^{2} \psi + n^{2} k^{2} \psi = 0$ by employing the boundary element method (BEM)~\cite{W03} under outgoing boundary conditions. This outgoing condition inherently induces non-Hermiticity by allowing energy to escape, leading to complex eigenvalues characteristic of open systems~\cite{W03, GC21}. In particular, we focus on transverse magnetic modes in a two-dimensional elliptical cavity in the $x$-$y$ plane, where $k$ is the wave number and $\psi$ is the $z$-component of the electric field.

\section{From Hermitian to Non-Hermitian Coupling: A Toy Model Approach}

We extend conventional Hermitian coupling to non-Hermitian coupling by introducing a $2\times2$ non-Hermitian Hamiltonian toy model. This model provides a simplified framework for capturing the essential features of non-Hermitian coupling in mode dynamics, offering physical insights beyond purely mathematical formulations. Following Ref.~\cite{HI14}, we define the real parts of $\lambda$ as functions of $\alpha$: $\lambda^{1}_{r}(\alpha) = 1 - \frac{\alpha}{2}$ and $\lambda^{2}_{r}(\alpha) = \sqrt{\alpha}$. The imaginary part $\lambda_{i}$ remains constant for analytical tractability. The non-Hermitian coupling term $g$ is chosen to model the rapid variation of coupling strength as the energy difference between the two interacting modes changes, inspired by the Landau-Zener effect, where transition probabilities between energy levels exhibit an exponential dependence:
\begin{equation}
    g(\alpha,\beta) = g_{c} \left[(1-\beta) + i\beta \right] \exp\left[- (\lambda^{1}(\alpha) - \lambda^{2}(\alpha))^2 \right],
\end{equation}
where $g_{c}$ is the coupling coefficient, determining the interaction strength between modes. The parameter $\beta$ controls the transition between Hermitian and non-Hermitian coupling regimes: when $\beta = 0$, the coupling remains purely real, corresponding to conventional Hermitian dynamics, whereas $\beta = 1$ results in purely imaginary coupling, leading to a fully non-Hermitian interaction. For intermediate values of $\beta$, the system experiences a gradual transition between these two regimes.

To analyze the transition from Hermitian to non-Hermitian coupling, we investigate two key phenomena:
\begin{enumerate}
    \item Landau-Zener avoided crossing (Fig.~\ref{Figure-1}a--c)~\cite{JE29,KS18},
    \item Width bifurcation associated with superradiance and subradiance (Fig.~\ref{Figure-1}d--f)~\cite{HI14,HI13}.
\end{enumerate}
The real parts of the eigenvalues, $\lambda_{r}$, exhibit an avoided crossing (Fig.~\ref{Figure-1}a), while the imaginary parts, $\lambda_{i}$, undergo a direct crossing (Fig.~\ref{Figure-1}b). The overlap integral of the two interacting eigenstates,
$O_{L}(i,j) = \frac{1}{X_{i}X_{j}} \iint dxdy |\psi^{*}_{i}(x,y)\psi_{j}(x,y)|,$ where $X_{i}=\sqrt{\int dxdy|\psi_{i}(x,y)|^{2}}$, reaches a maximum at $\chi \simeq 0.167$, indicating the presence of an \textsf{EP} (Fig.~\ref{Figure-1}c).

Width bifurcation, in contrast, exhibits different behavior. The real parts $\lambda_{r}$ cross and remain close over a finite range, while the imaginary parts $\lambda_{i}$ bifurcate at two critical points, with their relative difference maximized at the bifurcation center. This behavior corresponds to superradiance and subradiance in quantum optics~\cite{HI14,IJ15,Volya2005, Gross1982}. The overlap integral $O_{L}$ is maximized at the bifurcation edges (Fig.~\ref{Figure-1}f), suggesting the presence of two \textsf{EP}s. Additionally, this form effectively captures sudden changes in coupling as $\beta$ varies, particularly near \textsf{EP}s, which are characteristic of non-Hermitian systems.

These observations indicate that Landau-Zener interactions and width bifurcation arise from distinct physical mechanisms. However, a unified framework emerges when extending the system from Hermitian to non-Hermitian coupling. The parameter $\beta$ plays a crucial role in bridging the transition from Landau-Zener avoided crossing to width bifurcation. When $\beta = 0$, the coupling remains purely real, leading to a conventional Landau-Zener transition. As $\beta$ increases, a non-Hermitian coupling emerges, and the imaginary part of the eigenvalues $\lambda_i$ begins to bifurcate.
 At $\beta = 1$, the coupling is purely imaginary, and the system fully transitions into a width bifurcation regime. This smooth transition, controlled by $\beta$, highlights the fundamental connection between Landau-Zener interaction and non-Hermitian superradiance/subradiance effects.

To validate this framework, we compute the eigenvalues $\lambda_{\pm}$ using $g(\alpha,\beta)$ with initial values $g_{c} = 0.043$, $\lambda^{1}_{i} = 1.05$, and $\lambda^{2}_{i} = 1.07$, as shown in Fig.~\ref{Figure-1}g. The results confirm a smooth transition between the two regimes:
\begin{itemize}
    \item At $\beta = 0$, $g$ is purely real, reproducing the Landau-Zener interaction (Fig.~\ref{Figure-1}a, b).
    \item As $\beta$ increases to 0.8, the real parts $\lambda_{r}$ cross, and the imaginary parts $\lambda_{i}$ exhibit width bifurcation.
    \item At $\beta = 1$, $g$ becomes purely imaginary, and the system fully transitions to width bifurcation behavior (Fig.~\ref{Figure-1}d, e).
\end{itemize}
 Thus, unlike previous studies that treat Landau-Zener transitions and width bifurcation as distinct physical phenomena, our work presents a unified framework where these effects emerge as part of a continuous transition tuned by the parameter $\beta$.

\section{Avoided Crossing Transitions in Optical Microcavities and Toy Models}
 Determining the precise locations of \textsf{EP}s in microcavities, which are ubiquitous in open physical systems, is analytically challenging due to the intricate structure of the non-Hermitian Hamiltonian matrix elements. As a result, numerical methods become indispensable for identifying them. A particularly effective approach is to search for parameter regions where strong and weak interactions transition, as \textsf{EP}s manifest as singular points in these transitions~\cite{S08,WA90,KS18}. In such regions, the eigenvalues exhibit a distinctive transition: an avoided crossing in the real part ($\lambda_r$) accompanied by a crossing in the imaginary part ($\lambda_i$) transitions into an avoided crossing in $\lambda_i$ with a crossing in $\lambda_r$, and vice versa. This characteristic shift provides a definitive signature for identifying \textsf{EP}s with high precision.

\begin{figure}
\centering
\includegraphics[width=8.8cm]{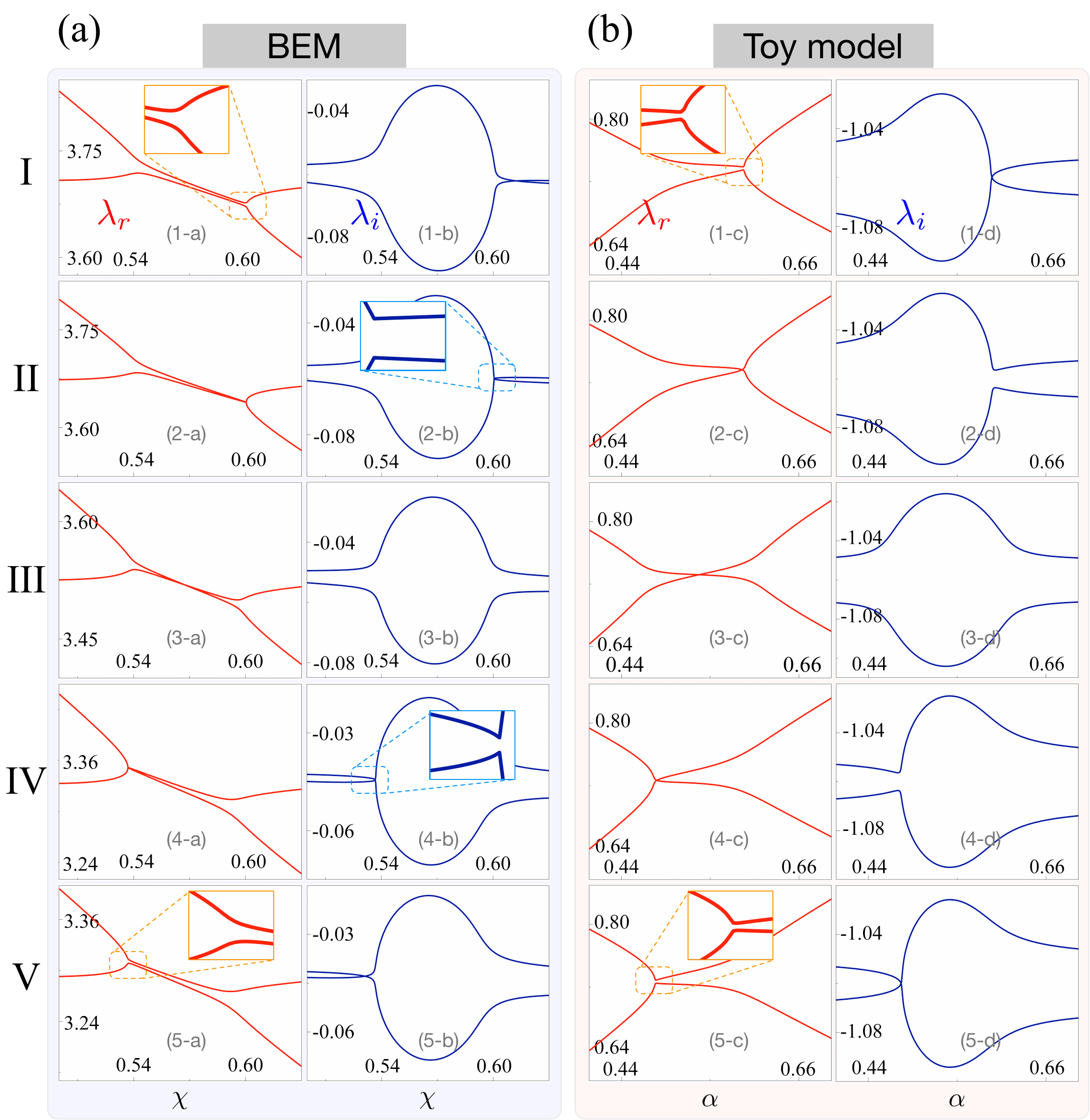}
\caption{Representative eigenvalue trajectories corresponding to Table~\ref{table-1}. Left panels: eigenvalues from our BEM simulation. Right panels: eigenvalues from the toy model Hamiltonian. Both panels exhibit similar trends.}
\label{Figure-2}
\end{figure}

\begin{table}
\centering
\renewcommand{\arraystretch}{1.5} 
\begin{tabular}{l|c|c|c|c|c|c|c} 
\hline\hline
{} & $n_{\tn{in}}$ & $g_{c}$ & $\beta$ & $\gamma_1$ & $\gamma_2$ & $\lambda_r$ & $\lambda_i$ \\ \hline
Class I & 2.6 & 0.043 & 0.76 & $1.05$ & $1.07$ & A.C. & C. \\ \hline
Class II & 2.6257 & 0.043 & 0.78 & $1.05$ & $1.07$ & C. & A.C. \\ \hline
Class III & 2.74 & 0.043 & 1 & $1.05$ ($1.07$) & $1.07$ ($1.05$) & C. & A.C. \\ \hline
Class IV & 2.9036 & 0.043 & 0.78 & $1.07$ & $1.05$ & C. & A.C. \\ \hline
Class V & 2.94 & 0.043 & 0.76 & $1.07$ & $1.05$ & A.C. & C. \\
\hline\hline
\end{tabular}
\caption{Representative parameters for successive transitions of avoided crossings in the microcavity and toy model. Class \text{I} exhibits an avoided crossing of $\lambda_{r}$ with a crossing of $\lambda_{i}$. Classes \text{II}, \text{III}, and \text{VI} exhibit an avoided crossing of $\lambda_{i}$ with a crossing of $\lambda_{r}$. Class \text{V} mirrors Class \text{I} with an avoided crossing of $\lambda_{r}$ and a crossing of $\lambda_{i}$.}
\label{table-1}
\end{table}

We identified five representative classes of avoided crossing transitions in both the BEM and toy models, as shown in Fig.~\ref{Figure-2}. The left panels display BEM simulation results, while the right panels show eigenvalue trajectories from the toy model for the parameters listed in Table~\ref{table-1}. In the BEM simulations, eigenvalues were obtained by varying the deformation parameter $\chi$ in the range $0.05\leq\chi\leq 0.63$ at each fixed $n_{\rm in}$. The numerical and theoretical results exhibit strong agreement across all classes.

\begin{itemize}
    \item Class \text{I}: Avoided crossing of $\lambda_r$ in (1-a), (1-c), and crossing of $\lambda_i$ in (1-b), (1-d).
    \item Classes \text{II}, \text{III}, \text{IV}: Avoided crossing of $\lambda_i$ in (2-b, 3-b, 4-b) and (2-d, 3-d, 4-d), with crossing of $\lambda_r$ in (2-a, 3-a, 4-a) and (2-c, 3-c, 4-c).
    \item Class \text{V}: Similar to Class \text{I}, with avoided crossing of $\lambda_r$ in (5-a), (5-c), and crossing of $\lambda_i$ in (5-b), (5-d).
\end{itemize}
A closer inspection of the BEM results reveals two critical transition points at $\chi\simeq 0.6$ and $\chi\simeq 0.53$, corresponding to a pair of \textsf{EP}s.

The dominant factor governing these transitions is the imaginary part of the coupling coefficient. Transitions occur when the imaginary component remains larger than the real component throughout the process, maintaining the width bifurcation structure, as shown in Fig.~\ref{Figure-1}(g). Another critical factor is the value of $\beta_c$, which governs transitions between these different classes of avoided crossings. In our examples, $\beta_c$ is approximately 0.765, with transitions occurring within the range $0.76\leq\beta\leq0.78$. When $\beta < \beta_c$ (Class \text{I}, Class \text{V}), the real part undergoes an avoided crossing, while the imaginary part exhibits a direct crossing. Conversely, for $\beta > \beta_c$ (Class \text{II}), this behavior is reversed. At $\beta=1$ (Class \text{III}), the system fully enters the width bifurcation regime, and as $\beta$ decreases back toward $\beta_c$ (Class \text{IV}), the transition gradually reverses, ultimately mirroring the Class \text{I} behavior in Class \text{V}. Additionally, the crossover position is influenced by the relative magnitude of the loss parameters $\gamma_i$. The transition occurs at $\alpha \approx 0.61$ when $\gamma_2 > \gamma_1$, whereas when $\gamma_1 > \gamma_2$, it shifts to $\alpha \approx 0.52$. This observation implies that dependence on loss asymmetry can also actively modify eigenvalue trajectories and mode coupling, offering a means for precise control over \textsf{EP} positioning.

\begin{figure*}
\centering
\includegraphics[width=\textwidth]{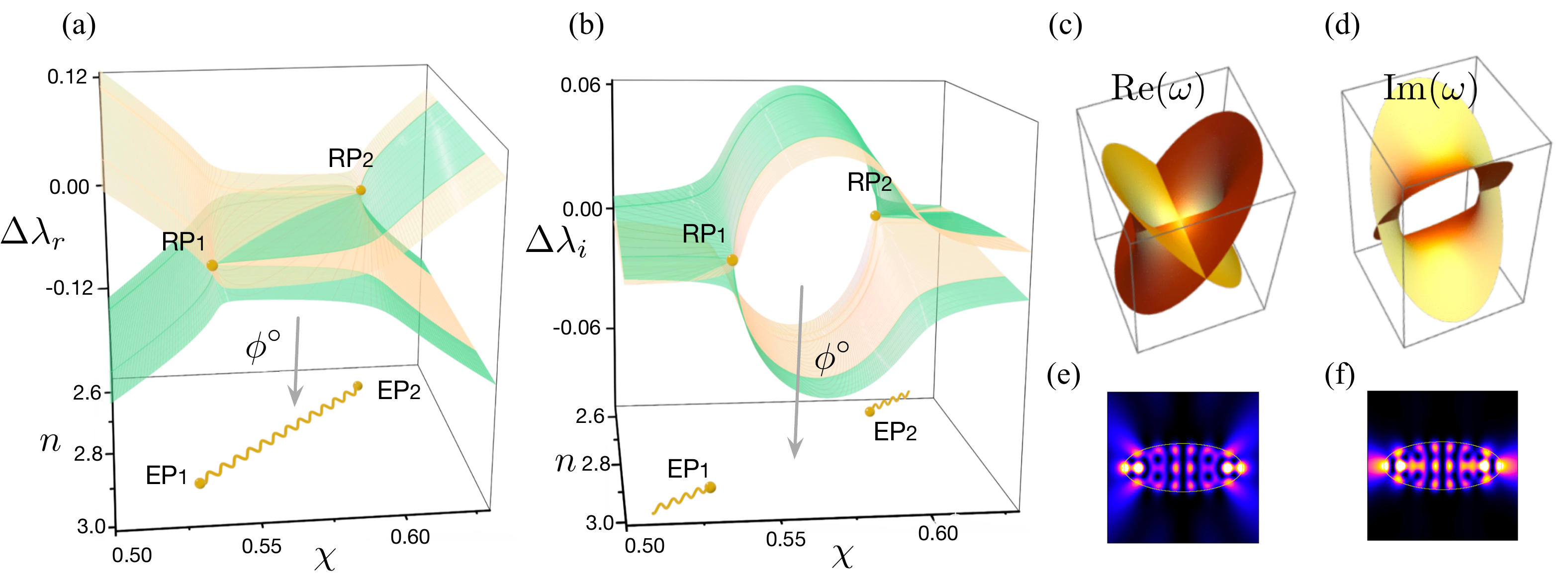}
\caption {Riemann surfaces (covering space) of the pair of \textsf{EP}s in an optical microcavity are displayed in the parameter plane (base space). The Riemann surface of the real part is shown in (a), while the imaginary part is presented in (b). Both surfaces are projected onto the base space through the covering map $\phi^{\circ}$. The pair of \textsf{EP}s is located at $n\simeq 2.6257, \chi=0.6001$ and $n\simeq 2.9036, \chi=0.5372$, respectively. Figures (c) and (d) were generated using \textit{Mathematica} with the function $f(z)=\sqrt{(z-z_{1})(z-z_{2})}=w$. Figures (e) and (f) are intensity plots of eigenfunctions at $\textsf{EP}_1$ and $\textsf{EP}_2$, respectively.}
\label{Figure-3}
\end{figure*}

\section{Riemann Surfaces and Exceptional Point Pairs in Optical Microcavities}
We analyzed the successive transitions associated with five representative avoided crossings in the range \(0.5\leq \chi \leq 0.63\) and \(2.6\leq n \leq 2.94\), identifying two critical points at approximately \(n\simeq 2.6, \chi=0.6\) and \(n\simeq 2.9, \chi=0.5\). The entire parameter space within this range was explored, leading to the construction of a complete eigenvalue structure that characterizes the eigenvalue topology.
The relative eigenvalue difference \(\Delta\lambda_{\pm} = \lambda_{\pm} - \lambda_{\rm AV}\) is introduced, where \(\lambda_{\rm AV} = (\lambda_{+}+\lambda_{-})/2\). This transformation eliminates global eigenvalue shifts and emphasizes the branching behavior near the \textsf{EP}s. The resulting Riemann surfaces of the real and imaginary parts of \(\Delta\lambda_{\pm}\) are shown in Fig.~\ref{Figure-3}(a) and (b), respectively. The \textsf{EP} pair is located at \((n\simeq 2.6257 ,\chi=0.6001)\) and \((n\simeq 2.9036 ,\chi=0.5372)\). The branch cut of the real-part Riemann surface $ Y_\tn{R}$ forms a direct curve connecting the two \textsf{EP}s, while that of the imaginary part, $ Y_\tn{I}$,  extends outward in opposite directions. This behavior deviates from conventional Riemann surfaces of the form \( f(z) = (z-z_{0})^{1/N} \), indicating a more intricate spectral topology.

A mathematical comparison of these surfaces is made using the function:
\begin{equation}
    f(z) = \sqrt{(z-z_{1})(z-z_{2})}=w, \quad \text{where} \quad z_1 \neq z_2.
\label{eq7}
\end{equation}
This equivalence is supported by Fig.~\ref{Figure-3}(c) and (d), which display Riemann surfaces generated using \textit{Mathematica} based on \( f(z)=w \). The structural similarity between numerical results in Fig.~\ref{Figure-3}(a) and (b) and these computed surfaces in Fig.~\ref{Figure-3}(c) and (d) validates this topological equivalence. Building upon previous studies, which primarily analyzed eigenvalue trajectories in isolated avoided crossings of superradiant and subradiant states~\cite{HI13, HI14, Persson00}, our work establishes a systematic framework for characterizing the global spectral topology of \textsf{EP} pair, revealing novel topological features that emerge beyond individual eigenvalue crossings.

For a rigorous mathematical framework, the parameter plane, defined as $\mathbb{P} = \{(n, \chi) : n, \chi \in \mathbb{R} \}$, is mapped to a complex plane, providing a natural setting to describe branching structures and multi-valued eigenfunctions. A homeomorphism $\rho: \mathbb{P} \to \mathbb{C}$ is established such that $\rho(\textsf{EP}_1) = z_1$ and $\rho(\textsf{EP}_2) = z_2$. This mapping associates \textsf{EP}s in the parameter space with branch points in the complex plane. The function $\rho$ induces a conformal structure on $\mathbb{P}$, which lifts to the Riemann surface $Y$, ensuring that the projection $\phi: Y \to \mathbb{P}$ is holomorphic, as shown in Fig.~\ref{Figure-3}.

Classical results in Riemann surface theory allow the projection $\phi$ to be expressed as a restriction of the coordinate projection $p_1: \mathbb{C} \times \mathbb{C} \to \mathbb{C}$, where $p_1(z, \omega) = z$. A polynomial $F(z, \omega)$ is introduced so that its zero locus represents the Riemann surface:
\begin{equation}
    \mathcal{Z}(F) = \{(z_0, \omega_0) \in \mathbb{C} \times \mathbb{C} \mid F(z_0, \omega_0) = 0 \}.
\end{equation}
Setting $F(z, \omega) = \omega^2 - (z - z_1)(z - z_2)$ with $z_1 = i$ and $z_2 = -i$, the branch points associated with the pair of \textsf{EP}s are identified. This formulation explicitly links the topological structure of the Riemann surface to the spectral behavior of the microcavity system.
Characterizing the \textsf{EP} locations in terms of critical points requires satisfying the condition $\partial_\omega F(z_0,\omega_0) = 2\omega_0 = 0$, which identifies critical (or bifurcation) points. This reveals that $\mathcal{Z}(F)$ branches along $z$ as $z$ varies. Consequently, the preimages of the \textsf{EP} pair under $\phi$ are found at $(i,0)$ and $(-i,0)$, where $\partial_z F \neq 0$, enabling the application of the Implicit Function Theorem. The local branching structure at $(i,0)$ and $(-i,0)$ is examined by introducing biholomorphic functions $\psi_+$ and $\psi_-$, which map small neighborhoods $D_+$ and $D_-$ of $0$ in the $\omega$-variable plane $\CC$ to neighborhoods of $i$ and $-i$ in the $z$-variable plane $\CC$, respectively, satisfying $F(\psi_\pm(\omega), \omega) = 0$ for all $\omega \in D_\pm$. Expanding \( z = \psi_\pm(\omega) \) in a Taylor series near \( \omega = 0 \), we obtain:
\begin{align}\label{localExpression}
\psi_\pm(\omega)=\pm i \mp \frac{i}{2}\omega^2+O(\omega^3).
\end{align}

Moreover, the relationship between Riemann surfaces and parameter space is rigorously examined using covering theory, which offers a structured framework for understanding their interplay, especially near \textsf{EP}s. A holomorphic covering map $\phi: Y \, (\text{Riemann surface}) \to X \, (\text{parameter space})$ relates these two spaces by mapping each ramification point $\fRP_i$ in $Y$ to its corresponding exceptional point $\textsf{EP}_i$ in $X$. Here, $\fRP_i$ denotes a point where $\phi$ fails to be a local homeomorphism. Removing these ramification points reduces the covering map to an unbranched covering, $\phi^\circ: Y^\circ \to X^\circ$, resulting in a smooth and well-defined map. Near each \textsf{EP}, the functions $\psi_{\pm}(\omega)$ act as local (branched) covering maps. From Eq.~\ref{localExpression}, the ramification index at each RP is $e_{\text{RP}} = 2$, indicating that the eigenvalues form a two-sheeted Riemann surface. Thus, a small variation in the system parameter leads to a characteristic square-root splitting of eigenvalues (see Sections 1--5 in the Supplementary Information for details).

\begin{figure*}
\centering
\includegraphics[width=17.0cm]{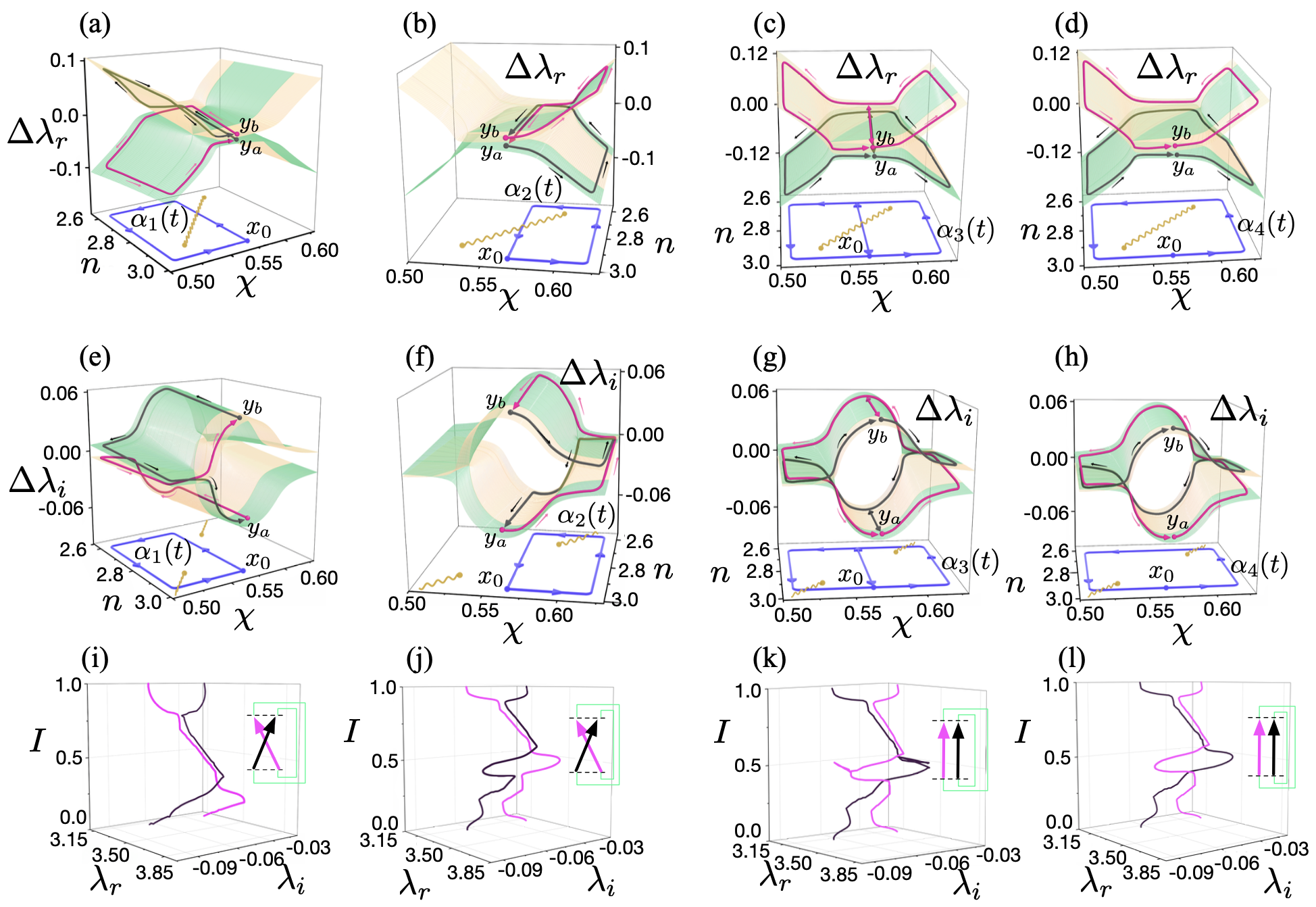}
\caption{
Path lifts on the Riemann surfaces (covering space) from the parameter space (base space) and the corresponding braids constructed from these lifts are illustrated. The base point of the control loop in the parameter space, denoted as $x_{0}$, corresponds to the fibers on the Riemann surfaces: $y_{a}$ (superradiance) and $y_{b}$ (subradiance).
Control loop $\alpha_{1}$ encircling $\textsf{EP}_{1}$ and its corresponding path lift on the Riemann surface: real part (a), imaginary part (e).
Control loop $\alpha_{2}$ encircling $\textsf{EP}_{2}$ and its corresponding path lift on the Riemann surface: real part (b), imaginary part (f).
Control loop $\alpha_{3}$ separately encircling $\textsf{EP}_{1}$ and $\textsf{EP}_{2}$, with its corresponding path lift on the Riemann surface: real part (c), imaginary part (g).
Control loop $\alpha_{4}$ encircling $\textsf{EP}_{1}$ and $\textsf{EP}_{2}$ simultaneously, with its corresponding path lift on the Riemann surface: real part (d), imaginary part (h). (i)--(l): Braids obtained from the path lifts for the control loops $\alpha_{1}$, $\alpha_{2}$, $\alpha_{3}$, and $\alpha_{4}$, respectively.
}
\label{Figure-4}
\end{figure*}

\section{Path Lifts and Braid Structures of Exceptional Point Pairs}
Since the control parameters of the system reside in the base space while the eigenvalues form Riemann surfaces in the covering space, fully characterizing eigenvalue topology requires lifting parameter space paths into the covering space. This path lifting provides a systematic framework for understanding eigenvalue transitions, particularly in the presence of \textsf{EP}s, where eigenvalues exhibit nontrivial topological behavior.

A systematic analysis of these lifted structures begins with the unique path lift theorem. Consider a (branched) covering map \( \phi:Y \to X \), where \( Y \) is the covering space and \( X \) is the base space. Given a path \( \alpha:I=[0,1] \to X^\circ \) that avoids branch points, for each \( p\in \phi^{-1}(\alpha(0)) \), there is  a unique lifted path \( \tilde{\alpha}: [0,1] \to Y^\circ \) such that \( \tilde{\alpha}(0)=p \) and \( \phi \circ \tilde{\alpha}=\alpha \). In our case, the covering spaces (Riemann surfaces) are designated as \( Y_\tn{R} \) for the real part and \( Y_\tn{I} \) for the imaginary part, both sharing a common base space \( X \) representing the parameter space.

\vspace{-0.5em}   
Applying this framework to our system, we examine the representative lifting of control loops onto the covering space. In Fig.~\ref{Figure-4}, four control loops \( \alpha_{i=1,2,3,4} \) are based at \( \alpha_i(0) = \alpha_i(1) = x_0 \in X \), and the   fibers over $x_0$ are given by \[\phi^{-1}(x_0) = \{y_a \text{ (superradiance)}, y_b \text{ (subradiance)}\}.\]
From these fibers, a total of 16 lifted paths emerge on the Riemann surfaces, as illustrated in Fig.~\ref{Figure-4}. Each \( \alpha_i \) has unique lifts \( \tilde{\alpha}^a_R(t) \) and \( \tilde{\alpha}^a_I(t) \) to \( Y_\tn{R} \) and \( Y_\tn{I} \), respectively, starting at \( y_a \). Likewise, there are unique lifts \( \tilde{\alpha}^b_R(t) \) and \( \tilde{\alpha}^b_I(t) \) to \( Y_\tn{R} \) and \( Y_\tn{I} \), respectively, starting at \( y_b \). Using these conventions, we define two parametric curves in \( \mathbb{C} \), as
\[ \lambda^a(t) = \tilde{\alpha}^a_R(t) + i \tilde{\alpha}^a_I(t) \in \mathbb{C} \text{ and }  \lambda^b(t) = \tilde{\alpha}^b_R(t) + i \tilde{\alpha}^b_I(t) \in \mathbb{C}. \]

\vspace{-0.5em}   
These curves naturally define a braid in the topological space \( S_{0,3}=\CC \setminus \{ \lambda_{\textsf{EP}_1 },\lambda_{\textsf{EP}_2}\} \), indicating that these curves represent two distinct points moving in \( S_{0,3} \). To rigorously classify these braids, we analyze their corresponding configuration space and fundamental group.
The configuration space \( \textnormal{UConf}_{2}(S_{0,3}) \) of unordered pairs of points in \( S_{0,3} \) is defined as follows:
\begin{align}
\textnormal{UConf}_{2}(S_{0,3}) = \{\{z, w\} \subset S_{0,3} \mid z \neq w\}. \big.
\end{align}

The fundamental group of this configuration space, which characterizes how two points can move and be continuously interchanged without colliding, defines the second braid group of \( S_{0,3} \):
\begin{align}
B_2(S_{0,3}) \doteq \pi_1\big(\textnormal{UConf}_{2}(S_{0,3})\big).
\end{align}
With this framework, we rigorously characterize the braid exchange process (see Sections 5-8 in the Supplementary Information for details). The function \( \beta: I \to \textnormal{UConf}_{2}(S_{0,3}) \), defined by \( t \mapsto (\lambda^a(t), \lambda^b(t)) \), represents a braid in \( B_2(S_{0,3}) \). As \( t \) varies from \( 0 \) to \( 1 \), the paths \( \lambda^a(t) \) and \( \lambda^b(t) \) continuously move in \( S_{0,3} \), leading to an exchange of two strands. This exchange can be understood by analyzing a loop in \( \textnormal{UConf}_{2}(S_{0,3}) \). Initially, at \( t = 0 \), the configuration is \( \beta(0) = (\lambda^a(0) = z_0, \lambda^b(0) = w_0) \). As \( t \) increases to \( 1 \), the endpoints swap positions, resulting in \( \beta(1) = (\lambda^a(1) = w_0, \lambda^b(1) = z_0) \). Expanding on this, we investigate how different encirclements of \textsf{EP}s influence the resulting braid structures. Specifically, $\alpha_1$ encircles $\textsf{EP}_1$ alone and corresponds to path lifts shown in (a) and (e), while $\alpha_2$ encloses only $\textsf{EP}_2$, with path lifts in (b) and (f). In contrast, $\alpha_3$ encircles both $\textsf{EP}_1$ and $\textsf{EP}_2$ separately, corresponding to (c) and (g), whereas $\alpha_4$ encloses them simultaneously, corresponding to (d) and (h).

\vspace{-0.1em}   
Figures~\ref{Figure-4}(i)--(l) reveal the effect of these encirclements on mode exchange. The braids associated with \( \alpha_1 \) and \( \alpha_2 \) exhibit mode exchange, while those for \( \alpha_3 \) and \( \alpha_4 \) do not. Notably, the braid for \( \alpha_3 \) [Fig.~\ref{Figure-4}(k)] is isotopic to that of \( \alpha_4 \) [Fig.~\ref{Figure-4}(l)], confirming the absence of mode exchange. More generally, different encircling paths yield identical braid structures as long as they enclose the same set of \textsf{EP}s. This dual possibility--where mode exchange occurs for some encirclements but not for others--is a fundamental characteristic of a robust switching mechanism, enabling precise and deterministic eigenstate transitions.

\setlength{\parskip}{10pt} 

\vspace{-\baselineskip}   
\section{Topological Switching of Q-Factor Driven by Exceptional Point Pairs}
The quality factor (Q-factor) measures the energy storage efficiency of a resonator and is defined as \(Q = \frac{\lambda_{r}}{2|\lambda_{i}|}\), where \(\lambda_{r}\) is the resonant wavelength and \(\lambda_{i}\) represents energy loss. In CW lasers, a higher Q-factor minimizes dissipation, sustains resonance, and reduces the threshold pump power \(P_{\text{th}}\) as \(P_{\text{th}} \propto Q^{-1}\). For pump power \(P_{\text{pump}}\) exceeding \(P_{\text{th}}\), the laser output follows \(P_{\text{out}} \propto \eta_{\text{slope}} \cdot (P_{\text{pump}} - P_{\text{th}})\), where \(\eta_{\text{slope}}\) represents the conversion efficiency. Thus, dynamic Q-factor control enables precise tuning of laser output power, enhancing performance across various applications~\cite{sil2004}.
\begin{figure*}
\centering
\includegraphics[width=\textwidth]{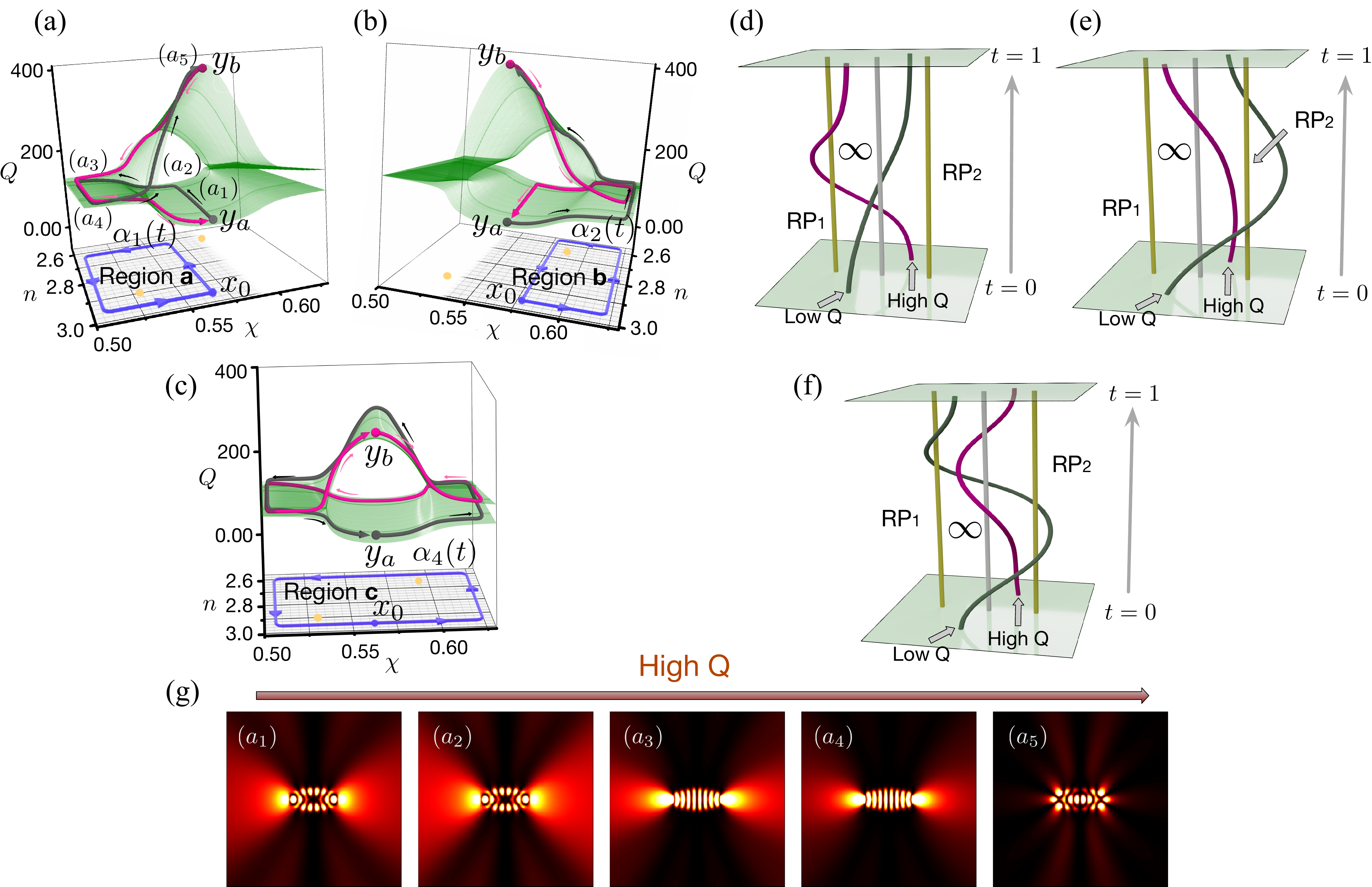}
\caption {Topologically protected switching of Q values facilitated by a pair of $\textsf{EP}$s. The base point ($x_{0}$) of the control loop in the parameter space corresponds to the fibers on the Riemann surfaces, $\{y_{a}$ (low Q), $y_{b}$ (high Q)$\}$. All loops homotopic to $\alpha_{1}$ (region \textbf{a}) in (a) or $\alpha_{2}$ (region \textbf{b}) in (b) result in Q-exchange, whereas loops homotopic to $\alpha_{4}$ (region \textbf{c}) in (c) do not induce Q-exchange.
Of particular importance, all isotopic braids corresponding to the strands (purple, dark gray) in (d) and (e) lead to Q-exchange, while any braids isotopic to the strands in (f) do not induce Q-exchange. This distinction underscores the relationship between topological braiding and Q-value transitions, which are determined by spatial regions and braiding configurations. In addition, Fig.~\ref{Figure-5}(g) shows a series of intensity plots of resonance modes from $a_{1}$ to $a_{5}$, as marked in Fig.~\ref{Figure-5}(a).}

\label{Figure-5}
\end{figure*}

\setlength{\parskip}{10pt} 

\vspace{-\baselineskip}   
To apply this framework to our specific system, we first examine the covering space \( Y_{Q} \) for the Q-factor. As shown in Fig.~\ref{Figure-4}, \( Y_{Q} \) is derived from the covering spaces \( Y_{R} \) and \( Y_{I} \), which are illustrated in Fig.~\ref{Figure-5}(a), (b), and (c).
Like $Y_\tn{I}$, the surface \( Y_{Q} \) is topologically equivalent (i.e. homeomorphic) to the plane with one point removed.
This structure therefore allows for systematic analysis of Q-value transitions, providing a robust framework for optimizing laser performance under varying conditions.

\setlength{\parskip}{10pt} 

\vspace{-\baselineskip}   
Unlike previous works on laser tuning, which primarily rely on gain-loss modulation or structural tuning for Q-factor control, our approach leverages topological braiding of \textsf{EP}s to achieve robust and deterministic Q-value transitions. This method overcomes the limitations of conventional designs, which are often sensitive to fabrication errors and parametric perturbations. This topological framework fundamentally constrains Q-value transitions to specific encirclements of \textsf{EP}s, dictated by the homotopy class of the control loops. As a result, Q-exchange occurs only when the system undergoes a nontrivial topological evolution, ensuring that only loops homotopic to \( \alpha_{1} \) in region \textbf{a} in (a) or \( \alpha_{2} \) in region \textbf{b} in (b), based at \( x_{0} \), induce an exchange of points in fibers $\{y_{a} \text{ (high Q)}, y_{b} \text{ (low Q)}\} \in Y_{Q}$. In contrast, no loop homotopic to \( \alpha_{4} \) in region \textbf{c} in (c) induces such an exchange, reinforcing that topological protection governs Q-factor switching rather than parametric variations alone.

\setlength{\parskip}{10pt} 

\vspace{-\baselineskip}   
This behavior can be formally established as follows. The braids representing Q-value switching are the lifted paths \( \tilde{\alpha}_i^a \) and \( \tilde{\alpha}_i^b \) of each \( \alpha_i \) to \( Y_Q \), originating at \( y_a \) and \( y_b \), respectively. As discussed in the previous section, this can be analyzed using the configuration space \( \textnormal{UConf}_{2}(Y_Q^\circ) \), consisting of unordered pairs of points in $Y_Q^\circ =\mathbb{C} \setminus \{\fRP_1, \fRP_2, \infty\} =S_{0,4}$ where each $\fRP_{i}$ is the (ramification) point in $Y_Q$, projected to $\textsf{EP}_{i}$. Note that the plane in Fig.~\ref{Figure-5}(d)--(f) represents the flattened $ Y_{Q}$, where the two intersection points with the yellow strands correspond to $\fRP_{1}$ and , $\fRP_{2}$ and the white strand marks the deleted point $\infty$. Then, the isotopy class of the curve defined by $t \mapsto (\tilde{\alpha}_i^a(t), \tilde{\alpha}_i^b(t))$ corresponds to a braid in
\begin{align}
B_2(Y_Q^\circ) = \pi_1\big(\textnormal{UConf}_{2}(Y_Q^\circ)\big).
\end{align}
By examining Fig.~\ref{Figure-5}(d)--(f) or (a)--(c), we observe that the Q-value exchange is independent of the orientation of the encircling \textsf{EP}s, indicating that chirality is not required for Q-value transitions. Thus, Q-values can be switched topologically, with all loops in regions \textbf{a} and \textbf{b} inducing Q-exchange regardless of specific paths, while those in region \textbf{c} do not. This topological behavior translates into robust and noise-resistant Q-value transitions. Clearly, as shown in Fig.~\ref{Figure-5}(g), the resonance modes evolve from $a_{1}$ to $a_{5}$, corresponding to a gradual increase in the Q-factor. Specifically, $a_{1}$ exhibits the lowest Q-factor, while $a_{5}$ attains the highest, demonstrating the controlled transition of optical modes within the microcavity.
\vspace{-0.1em}   
This topologically robust Q-value control is particularly advantageous in practical photonic systems, where stable and precise Q-factor tuning is crucial for performance optimization. As a result, our approach offers significant benefits for various laser applications requiring high stability and noise resistance. In particular, compact lasers often rely on direct modulation methods such as injection current tuning in semiconductor lasers~\cite{Russer82, Yamaoka21}, pump power control in optically pumped systems~\cite{Wang20, He23}, and temperature regulation~\cite{Huang24, Lai22}, as AOMs and VOAs are challenging to implement due to spatial constraints~\cite{Liu23, He21, Xiang23, Mourikis24}. However, these approaches suffer from thermal instabilities, high precision requirements, and sensitivity to noise, limiting their stability and practical feasibility.
\vspace{-1.1em}   

In contrast, our approach provides continuous and noise-resistant Q-value tuning, making it ideal for compact systems requiring precise and stable output control. By overcoming these limitations, our \textsf{EP}-based approach improves stability and performance while providing resilience against parametric perturbations in compact laser systems that demand highly controlled output power. This advancement benefits applications such as medical lasers for surgery~\cite{MN20, Schomacker1990}, miniaturized LiDAR for high-accuracy sensing~\cite{HuangX24, Jeong24}, atomic clocks for precision timekeeping~\cite{Zheng24, Hinkley13}, and optical tweezers for microscopic manipulation~\cite{Bustamante21, Grun24}. These results underscore the broad impact of topologically protected Q-factor switching in next-generation photonic systems.

\section{CONCLUSIONS}
We introduce a topologically protected Q-factor and mode-switching mechanism in an optical microcavity, leveraging paired exceptional points (\textsf{EP}s) governed by strong imaginary coupling and braid isotopy in a non-Hermitian two-level system. Simulations confirm that the high-Q/low-Q ratio is approximately 6 in the extreme low-frequency regime for numerical convenience, and at higher \( kr \), the high-Q/low-Q ratio is expected to increase further due to stronger mode confinement.
\vspace{-1.1em}   

Unlike chaotic microcavities, our simple elliptical design eliminates complex engineering, enabling accessible Q-factor transitions. Our results align with previous experimental observations of \textsf{EP}s in optical microcavities~\cite{Lee2009}, which demonstrated topological feature of an \textsf{EP} under realistic conditions. The proposed mechanism extends these findings by predicting novel behaviors that can be directly tested in tunable loss microcavities.
\vspace{-1.1em}   

This approach extends to broader systems, including topological semimetals, where the Riemann surface of an \textsf{EP} pair exhibits a topological equivalence to their band structures~\cite{Zhou2018, Su2021, Bergholtz2021}. Furthermore, eigenvalue braiding in this context is analogous to anyonic worldlines in topological quantum computation~\cite{Zhou2021, Bonesteel2005}. Our findings provide a theoretical foundation for advancing this line of research (Kyu-Won Park et al., in preparation). Furthermore, the observed superradiance and subradiance suggest potential applications in dissipative quantum optics. By bridging non-Hermitian and topological physics through \textsf{EP} physics, this framework enables topologically protected stable laser operation and advances quantum photonic applications.

\section{Acknowledgments}
This work was supported by the National Research Foundation of Korea (NRF) through a grant funded by the Ministry of Science and ICT (RS-2023-00211817, RS-2022-NR072395, NRF-2022M3H3A1098237), the Institute for Information \& Communications Technology Promotion (IITP) grant funded by the Korean government (MSIP) (No. 2019-0-00003; Research and Development of Core Technologies for Programming, Running, Implementing, and Validating of Fault-Tolerant Quantum Computing Systems), and Korea Institute of Science and Technology Information.

\bigskip
\section*{Disclosures}
The authors declare no conflicts of interest.

\section*{Author Contributions}
K.-W.P. proposed the study and performed theoretical analysis. K.K. performed the theoretical and mathematical analysis. K.J. supervised the overall investigation. K.-W.P., K.K., J.K., M.C., and K.J. wrote the manuscript. K.-W.P. and K.K. contributed equally to this work.

\section*{Supplementary Material: Branched Covering in Riemann Surfaces}

\subsection*{1. Covering Space and Covering Map}
A \textit{covering space} of a topological space $X$ is a topological space \(Y\) equipped with a continuous map \(\phi: Y \to X\) (called a \textit{covering map}) satisfying the following condition:
\begin{itemize}
    \item For every point \(x \in X\), there exists an open neighborhood \(U \subseteq X\) such that \(\phi^{-1}(U)\) is a disjoint union of open sets \(\{V_i\}_{i \in I}\) in \(Y\), where each \(\phi|_{V_i}: V_i \to U\) is a homeomorphism (i.e. $U$ is \emph{evenly covered}).
\end{itemize}
By the existence of an evenly covered neighborhood, we can see that the preimage \(\phi^{-1}(x)\) is a discrete set for all \(x \in X\).
Also, the covering map \(\phi\) is a \textit{local homeomorphism} since \(\phi^{-1}(U)\) can be expressed as multiple disjoint copies of \(U\).

\subsection*{2. Branched Covering}
Now, we review the definition of branched coverings only for Riemann surfaces, as the general theory is very technical.
Consult \cite{Fox} for the general theory of branched covering for topological spaces.

If we are only concerned with Riemann surfaces, we can define a \textit{branched covering space} of a Riemann surface $X$ as a Riemann surface $Y$ equipped with a holomorphic map $\phi: Y \rightarrow X$, since the axioms of a general branched covering are automatically implied by holomorphicity.
Unlike in (unbranched) coverings, there are some points in $X$, which have no evenly covered neighborhoods.
We call such points \emph{branch points}.
Namely, we can summarize this phenomenon as follows:
\begin{itemize}
    \item For any unbranched point \(x \in X\), there is a evenly covered neighborhood $U$ of $x$ and the preimage \(\phi^{-1}(x)\) consists of exactly \(d\) points, where \(d\) is the degree of the covering map.
    \item At any branch point \(x \in X\),
    there is no evenly covered neighborhood of $x$. In other words, the covering is not local homeomorphic at $x$.
\end{itemize}
\subsection*{3. Local Behavior of a Branched Covering}
To investigate the topological structure of a branched covering near branch points, we consider ramification points.
A point \(y \in Y\) is called a \emph{ramification point} if the derivative \(d\phi(y)\) at $y$ is $0$.
The ramification points map to branch points under \(\phi\).

By the defining condition, near a ramification point \( y \in Y \), the map \( \phi \) can be expressed locally as:
\[
    \phi(z) = \phi(y) + a_k (z - y)^k + \dots, \quad a_k \neq 0,
\]
where \( k = e_y \) is the \textit{ramification index} of \( y \). If \( e_y > 1 \), the point \( y \) is a non-trivial ramification point.
Otherwise, \( e_y = 1 \) and $\phi$ is a local homeomorphism near $y$.

This local expression shows that near \( y \), the map \( \phi \) behaves like the function \( z \mapsto z^k \). The ramification index \( k \) determines the number of ``sheets" of the covering space merging at the branch point \( x = \phi(y) \). For ramification points with \( e_y = k \), the preimage \( \phi^{-1}(x) \) will contain fewer distinct points due to this merging.


This structure can be better understood by considering the base point \( x \) in \( X \) rather than the lifted point \( y \) in \( Y \), as illustrated in the following commutative diagram:

\[
\xymatrix{
Y \ar[d]_\phi
& \phi^{-1}(\tilde{U}_x) \ar[r]^{\sim}_{\tilde{\varphi}} \ar[d]_\phi &  \bigsqcup_{y\in I_x}D_y \ar[d]^{\sqcup_y \phi_y} \\
X & \tilde{U}_x \ar[r]_{\tilde{\psi}}^\sim & D
}
\]

For an open neighborhood \( \tilde{U}_x \) of each point \( x \in X \), there exists a biholomorphic map \( \tilde{\psi}: \tilde{U}_x\longrightarrow D \) to some disk \( D \), satisfying \( \tilde{\psi}(x)=0 \). Additionally, we define another biholomorphic map \( \tilde{\varphi}: \phi^{-1}(\tilde{U}_x) \longrightarrow \bigsqcup_{y\in I_x} D_y \), where the disjoint union of disks is indexed by the set \( I_x = \phi^{-1}(x) \), and each map satisfies \( \tilde{\varphi}(y)=0 \in D_y \). Thus, we obtain, by taking $\tilde{\psi}\circ \phi \circ \tilde{\varphi}^{-1}$, the local expression of the covering map $\phi$:
\[
    \phi_{y}: D_{y} \longrightarrow D, \quad z \longmapsto z^{e_{y}}.
\]
This formulation provides a clear and structured representation of how the covering map behaves in the neighborhood of a branch point, emphasizing the local biholomorphic structure induced by the covering projection.

\subsection*{4. Example: \(\phi(z) = z^n\)}
Consider the map \(\phi: \mathbb{C} \to \mathbb{C}\), given by \(\phi(z) = z^n\):
\begin{itemize}
    \item \textbf{Branch Point:} The origin \(x = 0\) is a branch point because \(\phi^{-1}(0) = \{0\}\), while other points \(w \neq 0\) have \(n\) distinct preimages \(w^{1/n}\).
    \item \textbf{Ramification Point:} The origin \(z = 0\) is a ramification point with ramification index \(e_0 = n\).
\end{itemize}

To further understand this example, observe the behavior of the derivative:
\[
    d\phi(z) = n z^{n-1}.
\]
At \(z = 0\), we have \(d\phi(0) = 0\), indicating that the map \(\phi\) is not locally invertible at this point. Near \(z = 0\), the map \(\phi(z) = z^n\) ``wraps" the domain around the origin \(n\)-times, resulting in a single branch point at the origin \(x = 0\).
For any point \(w \neq 0\), the preimages \(w^{1/n}\) are distinct, showing that there are no branch points or ramification points away from the origin.

\subsection*{5. Relationship Between Branched and Unbranched Coverings}

\subsubsection*{1. Branched Covering Structure}
Let  \(\phi: Y \rightarrow X\) a proper holomorphic map  between two Riemann surfaces:
\begin{itemize}
    \item \(Y\): the branched covering space.
    \item \(X\): the base space onto which \(Y\) is mapped.
\end{itemize}

Now, we analyze its local homeomorphic properties. The collection \(S_{\phi}\) of all points in \(Y\) at which \(\phi\) fails to be a local homeomorphism is a discrete and closed subset of \(Y\). By the definition, \(S_{\phi}\) is the collection of non-trivial ramification points (\(\textsf{RP}\)).

\subsubsection*{2. Unbranched Covering as a Restriction}
By removing the branch points and their preimages, we obtain an unbranched covering:

\[
    \phi^\circ: Y^\circ = Y \setminus \phi^{-1}(\phi(S_{\phi})) \to X^\circ = X \setminus \phi(S_{\phi}).
\]
Here, \(\phi^\circ\) defines an ordinary covering map. In other words, outside the ramification locus,
$\phi$ behaves as a usual covering.

\subsubsection*{3. Application to Our Branched Covering}
Recall the covering $\phi:\cZ(F)\to \CC$ where $\cZ(F)$ is the zero locus of \[
F(z,\omega)=\omega^2-(z-z_1)(z-z_2)
\]
with $z_1\neq z_2$.
In this specific case, we can characterize the ramification set as:
\[
    S_{\phi} = \{\textsf{RP}_1, \textsf{RP}_2\}.
\]

Under the map \(\phi\), these ramification points project onto corresponding \textit{exceptional points} (\(\textsf{EP}\)) in the base space \(X\):

\[
    \phi(\textsf{RP}_i) = \textsf{EP}_i.
\]

Thus, the branch points in \(X\) arise precisely from the projection of ramification points in \(Y\), confirming the branched covering structure. This perspective provides a strict mathematical framework for understanding how branched coverings generalize classical unbranched coverings, ensuring a consistent interpretation of Riemann surfaces with singular points.

\subsection*{6. Unique Path Lifting Property and Braid Structure}
An important property of covering maps is the \textit{unique path lifting property}, which ensures that any path in the base space \(X\) can be uniquely lifted to a path in the covering space \(Y\), provided the starting point of the lift is specified.

Let \(\phi: Y \to X\) be a (branched) covering map, \(\alpha: [0, 1] \to X\) be any path in the base space avoiding the branch points \(\phi(S_{\phi})\), namely, $\alpha([0,1])\subset X^\circ$. For each point \(p \in \phi^{-1}(\alpha(0))\), there exists a unique path \(\tilde{\alpha}: [0, 1] \to Y^\circ\) such that:
\begin{itemize}
    \item \(\tilde{\alpha}(0) = p\), and
    \item \(\phi \circ \tilde{\alpha} = \alpha\),
\end{itemize}
that is, the following diagram commutes:
\[
\xymatrix{
 &  Y \ar[d]^{\phi} \\
I \ar[r]_{\alpha} \ar@{..>}[ru]^{\tilde{\alpha}} & X
}
\]
This property is critical to understand the relationship between paths on Riemann surfaces and their braid structure. Specifically, the lifting property ensures that each loop in the base space \(X\) correspond to a braid in the covering space \(Y\), with branch points introducing twists between the sheets of the covering.

\subsection*{7. Fundamental Group and Configuration Space}

\subsubsection*{Fundamental Groups}
Given a topological space $X$ and a point $x_0$ in $X$, we can define the \emph{fundamental group} \(\pi_1(X, x_0)\) which provides a topological invariant, characterizing the global structure of \(X\).
It is defined as the set of equivalence classes $[\gamma]$ of loops $\gamma$ based at a point \(x_0 \in X\), where two loops are considered equivalent if one can be continuously deformed into the other.

\vspace{2ex}
\noindent
\textbf{Definition:}
\[
\pi_1(X, x_0) = \{ [\gamma] : \gamma : [0, 1] \to X, \gamma(0) = \gamma(1) = x_0 \},
\]
where $[\gamma]$ denotes the collection of all paths that are equivalent to $\gamma$.

\vspace{2ex}
\noindent
\textbf{Key Properties:}
\begin{itemize}
    \item \textbf{Path-Connected Spaces:} For a path-connected space, \(\pi_1(X, x_0)\) does not depend on the choice of \(x_0\).
    \item \textbf{Simply Connected Spaces:} If every loop in \(X\) can be contracted to the base point (considered as a constant path), then \(\pi_1(X) = \{e\}\) where $e$ is the equivalence class of the constant path at the base point.
    Namely, $\pi_1(X)$ is the trivial group.
    \item \textbf{Product Spaces:} If \(X = X_1 \times X_2\), then \[\pi_1(X) \cong \pi_1(X_1) \times \pi_1(X_2).\]
\end{itemize}

\vspace{2ex}
\noindent
\textbf{Examples:}
\begin{itemize}
    \item \(\mathbb{C}\): The complex plane is simply connected, that is, \(\pi_1(\mathbb{C}) = \{e\}\).
    \item \(\mathbb{S}^1\): The circle has \(\pi_1(\mathbb{S}^1) \cong \mathbb{Z}\), where each equivalence class of loops corresponds to a unique integer, which represents the winding number.
\end{itemize}
Consult \cite{Hatcher} for a nice exposition about the general theory of fundamental groups.

\subsubsection*{Configuration Space}

The configuration space describes the possible arrangements of \(n\) distinct points within a topological space \(T\). It is a fundamental concept in mathematics and physics, particularly in the study of particle systems, motion planning, and braid groups.

\vspace{2ex}
\noindent
\textbf{Ordered Configuration Space:}
\[
F(T, n) = \{ (t_1, t_2, \dots, t_n) \in T^n \mid t_i \neq t_j \text{ for all } i \neq j \}.
\]
Here, each tuple \((t_1, t_2, \dots, t_n)\) represents the positions of \(n\) distinct points in \(T\), which are ordered by $1,\dots,n$.
Note that the space \(F(T, n)\) is a subset of \(T^n\) where no two points coincide.

\vspace{2ex}
\noindent
\textbf{Unordered Configuration Space:}
$C(T,n)$ is defined as a collection of subsets $A$ of $T$ that contains exactly $n$ distinct points of $T$.
Namely,
\[
C(T,n)=\{A\subset T: |A|=n\}
\]
where $|A|$ denotes the number of elements of $A$.
Unlike an element of $F(T,n)$, an element of $C(T,n)$ represents $n$ distinct position without ordering (or labeling).

\vspace{2ex}
\noindent
\textbf{Properties of Configuration Spaces:}
\begin{itemize}
    \item \textbf{Path-Connectedness:} If \(T\) is path-connected, then \(F(T, n)\) is typically also path-connected.
    \item \textbf{Dimensionality:} When $T$ is a manifold, the space \(F(T, n)\) is an open subset of \(T^n\), and thus has dimension \(n \cdot \dim(T)\).
    \item \textbf{Topology of avoidance:} The condition \(t_i \neq t_j\) ensures that points do not collide, which is important in applications such as robotics and fluid dynamics, where singular configurations (collisions) must be avoided.
\end{itemize}

\subsubsection*{Braid groups}
\begin{itemize}
    \item The fundamental group of the ordered configuration space, \(\pi_1(F(T, n))\), is the \textbf{pure braid group} \(P_n\) of $T$.
    Each element of $P_n$ is called a $n^{th}$-\emph{pure braid} on $T$.
    \item The fundamental group of the unordered configuration space, \(\pi_1(C(T, n))\) of $T$, is the \textbf{braid group} \(B_n\).
    Each element of $B_n$ is called a $n^{th}$-\emph{braid} on $T$
\end{itemize}


\subsection*{8. Braid Groups}
Given a topological space $T$, we can think of an element in $\pi_1(T,t_0)$ as the trajectory of a single point, starting at the base point $t_0$	
  and ending at $t_0$.
Under this interpretation, we can consider an
$n^{th}$-pure braid on $T$ as the trajectory of $n$ distinct points, where each point returns to its starting position without colliding with other points.
Similarly, an
$n^{th}$ braid on $T$ is the trajectory of $n$ distinct points, where the trajectory of each point either returns to its starting position or ends at the starting position of another point.
In both cases, the trajectory of  each point is called a \emph{strand} of a braid.

In fact, there are various version of interpretations of braids.
See \cite{Birman} for a general theory of braid groups.




\subsubsection*{Generators and Relations}
The braid group \(B_n\) is generated by \((n-1)\) standard generators \(\sigma_1, \sigma_2, \dots, \sigma_{n-1}\), where:
\begin{itemize}
    \item \(\sigma_i\) represents the crossing of the \(i\)-th strand over the \((i+1)\)-th strand.
    \item The group satisfies the following relations:
    \[
    \sigma_i \sigma_j = \sigma_j \sigma_i, \quad \text{if } |i-j| > 1,
    \]
    \[
    \sigma_i \sigma_{i+1} \sigma_i = \sigma_{i+1} \sigma_i \sigma_{i+1}.
    \]
\end{itemize}

\subsubsection*{Key Properties}
\begin{itemize}
    \item \textbf{Homomorphism to Symmetric Group:}
    The braid group \(B_n\) admits a natural surjective homomorphism $f$ to the symmetric group \(S_n\) on $\{1,2,\dots, n\}$, where each generator \(\sigma_i\) corresponds to the transposition swapping \(i\) and \(i+1\).
    \item \textbf{Relationship with Pure Braid Group:} The kernel of this homomorphism $f$ is \(P_n\), the pure braid group, which tracks braids without permutation of end points of strands.
    \item \textbf{Applications:} Braid groups play a crucial role in mathematical physics, knot theory, and algebraic topology. These appear in knot theory, algebra, and quantum computation, where the topology of particle paths and their equivalence is of interest.
\end{itemize}

\subsubsection*{Examples}
\begin{itemize}
    \item \(B_2\): The simplest braid group, with one generator \(\sigma_1\). It is isomorphic to \(\mathbb{Z}\), as any braid can be represented by integer powers of \(\sigma_1\).
    \item \(B_3\): Contains two generators \(\sigma_1\) and \(\sigma_2\), subject to the relations:
    \[
    \sigma_1 \sigma_2 \sigma_1 = \sigma_2 \sigma_1 \sigma_2.
    \]

\end{itemize}

\subsection*{Linking Path Liftings to Braids: A Mathematical Explanation}

The relationship between paths on the base space \(X\) of a branched covering \(\phi: Y \to X\) and the braid structure on $Y$ arises naturally through the \textit{unique path lifting property} and the topology of the configuration space. Here, we explain this connection step by step.

\subsubsection*{1. Path Lifting and Covering Spaces}
\begin{itemize}
    \item \textbf{Base Space \(X\)}: Let \(\alpha: [0, 1] \to X^\circ\) be a path that avoids branch points (\(\phi(S_\phi)\)).
    \item \textbf{Covering Space \(Y\)}: The unique path lifting property ensures that, for a point \(y_0 \in \phi^{-1}(\alpha(0))\), there exists a unique path \(\tilde{\alpha}: [0, 1] \to Y^\circ\) such that:
\[
\phi(\tilde{\alpha}(t)) = \alpha(t), \quad \tilde{\alpha}(0) = y_0.
\]
This lifted path \(\tilde{\alpha}\) lies in the covering space \(Y\), representing how the sheets of the covering ``follow" the path \(\alpha\).
\end{itemize}

\subsubsection*{2. Braids and Path Liftings}
\begin{itemize}
    \item  \textbf{Braids}: Set $B=\phi^{-1}(\alpha(0))$ and $n=|B|$.
    Consider the $n^{th}$-braid group on $Y^\circ$.
    Recall that $B_n(Y^\circ)$ is the fundamental group of $C(Y^\circ, n)$.
    For our purpose, we now assume that $B$ is the base point for $B_n(Y^\circ)$, that is, $B_n(Y^\circ)=\pi_1(C(Y^\circ, n), B)$.
    \item  \textbf{Strands as Lifted Paths}:
    The $n$ distinct lifting of $\alpha$ on $Y$ provides  a braid with \(n\) strands on $Y^\circ$.
    Each strand shows how a point in the covering space moves as the base path \(\alpha\) moves around.
\end{itemize}



\subsubsection*{3. Connection Between Path Lifting and Configuration Space}
While the braid group \(B_n = \pi_1(C(Y^\circ, n))\) is defined on the configuration space of \(Y^\circ\), the covering space \(Y\) provides a geometric way to visualize these braids. The lifted paths in \(Y\) align with the configuration space structure because:
\begin{itemize}
    \item Loops in \(C(Y^\circ, n)\) describe how points in \(Y^\circ\) exchange positions.
    \item The lifted paths in \(Y\) encode these exchanges as crossings and twists, corresponding to the braid structure.
\end{itemize}


\subsubsection*{4. Key Clarification and Conclusion}
By using \(Y^\circ\) as the topological space governing the motion of points, the braid structure can be naturally derived:
\begin{itemize}
    \item  \textbf{Base Space \(X\)}: The space where variables are defined, including branch points and the loop \(\alpha\).
    \item  \textbf{Covering Space \(Y\)}: The space where lifted paths form the braid structure.
    \item \textbf{Fiber}: The \(n\) points in the covering space \(Y\) corresponding to a single point in \(X\), which become the strands of the braid.
\end{itemize}

This interpretation ensures consistency between the discussion of EPs and the general definition of braids, seamlessly linking these concepts.
Also, see \cite{GoncalvesGuaschi} for the study about the relation between the braid groups of a surface and of its unbranched covering.

\end{document}

%% file: header_labeling.tex
\theoremstyle{plain}

\theoremstyle{definition}
\newtheorem{defn/}[equation]{Definition}

\theoremstyle{remark}
\newtheorem{rmk/}[equation]{Remark}
\newtheorem*{rmk*}{Remark}

\newtheorem*{claim*}{Claim}



%% file: header_newcommand.tex
\newcommand{\CC}{\mathbb{C}}

\newcommand{\cZ}{\mathcal{Z}}

